**Title –** Effect of thermal and cryogenic conditioning on flexural behavior of thermally shocked $Cu-Al_2O_3$ micro- and nano-composites.

**Authors –** Khushbu Dash, Sujata Panda and Bankim Chandra Ray.

**Author affiliations –**

**Corresponding author**

Khushbu Dash

Research Scholar

Dept. of Metallurgical and Materials Engg.

National Institute of Technology,

Rourkela – 769008

Phone – +91-9439281130(M)

Fax - +91-661-2465999

E-mail – khushbudash@gmail.com

**Co-authors**

Sujata Panda

M.Tech Scholar

Dept. of Metallurgical and Materials Engg.

National Institute of Technology,

Rourkela – 769008

Phone – +91-9861238102(M)

Fax - +91-661-2465999

E-mail – suj.panda@gmail.com

Bankim Chandra Ray

Professor

Dept. of Metallurgical and Materials Engg.

National Institute of Technology,

Rourkela – 769008

Phone – +91-9437221560 (M)

Fax - +91-661-2465999

E-mail – drbcray@gmail.com

# Effect of Thermal and Cryogenic Conditioning on Flexural Behavior of Thermally Shocked Cu-$Al_2O_3$ Micro- And Nano-Composites

**K. Dash*, S. Panda and B.C. Ray**

**Department of Metallurgical and Materials Engineering**

**National Institute of Technology Rourkela, Rourkela – 769008, India**

## Abstract

This investigation has used flexural test to explore the effect of thermal treatments i.e. high temperature and cryogenic environments on the mechanical property of $Al_2O_3$ particulate reinforced Cu metal matrix micro- and nano-composites in ex-situ and in-situ conditions. Cu-5 vol. % $Al_2O_3$ micro (10μm)- and nano (<50 nm)-composites fabricated by powder metallurgy route were subjected to up-thermal shock cycle (193 K to 353 K) and down-thermal shock cycle (353 K to 193 K) for different time periods followed by 3-point bend test. One batch of specimens (micro- and nano-composites) was conditioned at 353 K temperature and 193 K temperature separately followed by 3-point flexural test. High temperature flexural test was performed at 373 K and 523 K temperature on the micro- and nano-composites. All the fractured samples obtained after various thermal treatments were studied under scanning electron microscope (SEM). The development of thermal stresses quite often results in concentration of residual stresses at the particle/matrix interface eventually weakening it. Enhancement of flexural strength was recorded for down- as well as for up-thermal shock in microcomposites. The high temperature flexural strength of micro- and nano-composites is lower than at ambient temperature. The amelioration and declination in mechanical properties as a consequence of thermal shock, thermal conditioning and high temperature flexural testing have been discussed in the light of fractography.

*Corresponding author:

E-mail: khushbudash@gmail.com

Phone no: +91 661 2640369

## I. INTRODUCTION

Metal matrix composites (MMCs) are excellent candidates for structural components in the aerospace and automotive industries due to their high specific modulus, strength, and thermal stability. [1] Metal matrix composites show a combination of mechanical stiffness (reinforcement), a relatively low density and a high damping capacity (matrix). This combination makes it an attractive material for aerospace and relatively high temperature applications.

Strengthening of MMCs depends on the character and architecture of its constituents: the matrix and the reinforcement.

Differential thermal expansion is a prime cause of development of thermal stresses in composite materials. The thermal stresses generated depend on the reinforcement volume fraction, reinforcement geometry, and thermal expansion coefficient mismatch. [2-3] Most reinforcements are elastic up to their fracture point, while the matrix undergoes plastic deformation. [4] Processing induced reactions and thermal stresses can cause changes in the matrix microstructure. These microstructural changes in the matrix in turn can affect the mechanical and physical behavior of the composite. The mechanical properties of the composite depend on the relaxation mechanism of these thermal stresses. When the thermal stresses accumulated are high enough to exceed the elastic limit of the matrix, dislocations are produced in the vicinity of the reinforcement. [5-7]

A significant thermal expansion mismatch may result in de-cohesion at the particle/matrix interface and/or a possible matrix cracking, particle fragmentation due to thermal stress. [8] Such a concentration of thermal stresses around defects (second phase particles) often results in catastrophic failure of the composite. [9] Zhou et al. [9] reported that the initial crack occurs in the notched-tip region, wherein the initial crack is induced by void nucleation, growth and subsequent coalescence in the matrix materials or interface separation.

Improvements of fracture toughness are also ascribed to the mechanisms of crack blunting, shielding and crack deflection accompanied by particle pull out. The performance of particle reinforced composite is often controlled by the adhesion theories at the particle/matrix mating surfaces. Thermal expansion coefficient of metals are substantially greater compared to ceramics which leads to either enhancement or degradation of proximal contact between particle and matrix occurs under the influence of temperature gradient. [10] The 3-point flexural test results may reflect the tendency of the bond strength where the bonding level only is a variable. A need probably exists for an assessment of mechanical performance of such composites under the influence of thermal shock. Thermal stresses caused by temperature gradient should be given special attention in many application areas. A better understanding of interfacial properties and assessment of interfacial strength can help in evaluating the mechanical behavior of particle reinforced metal matrix composite materials. It is important to experimentally assess the degree of degradative effect of thermal conditioning during exposure to above-zero and sub-zero temperatures. Ma et al. [11] studied the cryogenic properties and fracture behavior of Al composites predicting the rise in tensile strength at cryogenic temperatures from the room temperature. They have also illustrated different fracture modes and features in cryogenic temperature range. Poza et al. [12] investigated the fracture mechanisms of Al composites at cryogenic and elevated temperatures. The fracture characters at elevated temperatures are dominated by interfacial de-cohesion rather than the reinforcement particle fracture.

High thermal conductivity of copper makes it a potent candidate for heat conduction applications in aerospace and automobile components. The high conduction coupled with high strength would complete the requirements for a high conduction-high strength material. These components often suffer high temperature on one surface and cryogenic environment on the other end. [13] The copper-alumina combination needs to be understood for the super critical applications in the aerospace industry. There is a paucity of literature in this field hence this investigation would lead to a comprehensive study of the composite with particle size variation. The advancement of science and technology has been rapidly demanding newer material which can endure extreme weathering exposures and excursion. This may necessitate the design of experimental process and procedures to generate data and findings which would lead to the prediction of reliability of mechanical performance of material behavior in unpredictably harsh and hostile environments. In this investigation we have subjected the Cu-$Al_2O_3$ micro- and nano-composites to different thermal treatment modules which consist of thermal shock treatment, thermal conditioning and high temperature flexural testing.

Boccacini et al. [14] have shown that thermal cycling and thermal fatigue in MMCs (in Mg-Al alloy system) leads to decrease in elastic modulus and density, these thermal stresses can be relieved by plastic deformation of matrix, cracking, debonding at interface followed by void formation. The Al alloy–SiC composite show cracks in response to repeated thermal cycling, the microhardness decreases with increase in number of thermal cycles. The interfacial cracks lead to a decrease in mechanical performance. The matrix-particle interfacial decohesion plays a major role in the detoriation of mechanical property than the damage accumulation. [15] Al/SiC p composites subjected to thermal cycling and its behavior was studied. Complex dislocation interactions take place during thermal cycling. [16] Dislocation debris and curved dislocation segments near particle matrix interface indicate relaxation.

Magnesium alloy reinforced with carbon fibers were fabricated via gas pressure infiltration technique and had been subjected to thermal shocks of +100 to -100°C. The number of thermal cycles has been varied from 1, 2, 5, 10, 20, 50 to 100 cycles followed by flexural tests. Bending modulus, ILSS values and microhardness values were studied after thermal cycling. Variation in microhardness symbolizes work hardening, viscolelastic recovery or aging. Thermal cycling led to mechanical damage via microcracks, microvoids and fiber protrusion by interfacial sliding.[17]

## II. EXPERIMENTAL

### A. *Fabrication of composites*

Copper powder (Loba Chemie) (average particle size~11μm, purity- >99.7%) (trace impurities: P, Sb, Fe, Pb, Mn, Ag, Sn) was used as the matrix material. Alumina powders (Sigma Aldrich) (average particle size~10μm and < 50 nm) were selected as the reinforcement material. The Cu–5% $Al_2O_3$ (10 μm) microcomposites and Cu–5% $Al_2O_3$ (< 50 nm) nanocomposites powders were blended separately. The specimens having dimensions (31.5 x12.7 x 6.3 $mm^3$) were prepared by compacting the powders at a pressure of 500 MPa as per ASTM B 925-08 for 3-point flexural test. The compacted specimens were sintered conventionally at 1173 K (900˚C) for 90 minutes in argon atmosphere. Three specimens were tested at each point of experimental condition under flexural loading. The average of two and/or three closer values of flexural test data was chosen for indicating the nature of variation.

### B. *Ex-situ thermal treatment*

The 3-point flexural test standard specimens were subjected to thermal shock environment with a 433 K (160˚C) temperature gradient by two separate routes. Thermal shock treatment of the micro- and nano-composites was done from 353 K (+80˚C) temperature to 193 K (-80˚C) temperature (down-thermal shock) for one batch of specimens and in the reverse order (up-thermal shock) for another batch of specimens in three different modules explained in Table. 1. After each thermal shock treatment, 3-point flexural test of each sample was performed immediately in an universal testing machine (Instron-5967) at a cross-head speed of 0.5 mm/min maintaining a span length of 26 mm. Choice of temperatures between +80˚ C and -80˚C was made to promote accelerated weathering which readily induces significant scale of damage and development in shorter span of time, this accelerated weathering data may be used to predict long term durability for application at higher temperatures than -80˚C. The same is true for +80˚C**.** In order to treat the samples for thermal conditioning the samples were categorized into 3 groups. The first group was treated at 353 K (+80˚C) in muffle furnace for 60 minutes isothermally. The blower of the furnace was on function for heat circulation and uniform heating of the samples. The second set of samples were treated at 193 K (-80˚C) in an ultra low temperature chamber for 60 minutes by isothermal holding. The third batch of samples was maintained at ambient temperature. 3-point flexural test was conducted immediately at room temperature after each thermal conditioning treatment. The loading rate and span length was maintained at 0.5 mm/min and 26mm respectively for all the 3-point bend tests.

C. *High temperature flexural testing*

High temperature 3-point flexural test was carried out at a temperature of 373 K (100˚C), 523 K (250˚C) on the micro- and nano-composites. The samples were kept inside the furnace chamber and the furnace was allowed to reach the required temperature. The 3-point flexural test was conducted just after the temperature attainment.

D. *SEM analysis*

The fracture surface of the all 3-point flexural test samples were observed under scanning electron microscopy (JEOL 6480 LV). The fractographic studies revealed the various possible fracture modes operating during the thermal shock, thermal conditioning, high temperature flexural test and also at ambient temperature.

## III. RESULTS AND DISCUSSION

The densification of Cu-5% $Al_2O_3$ microcomposite is 91.04% of theoretical density whereas the densification of Cu-5% $Al_2O_3$ nanocomposite is 85.2% of theoretical density measured by Archimedes method. The microstructures of Cu-5% $Al_2O_3$ microcomposites before and after down thermal shock treatment have been illustrated in Fig. 1. The fabricated microcomposites show good dispersion of alumina particles in copper matrix, whereas the down-thermal shock treated microcomposites show particle cracking and particle pull out too. Fig. 2 shows micrographs of Cu-5% $Al_2O_3$ nanocomposites before and after down-thermal shock treatment. The nanocomposite before treatment has been marked with alumina nanoparticles which show almost uniform distribution. The microstructure of nanocomposite after down-thermal shock shows pulled out agglomerated particle resting on the matrix. Figs. 1 & 2 demonstrate the reasonably visible uniform distribution of pores throughout the composite system. The objective of this investigation is to assess the variation in mechanical property with the thermal treatments, but not so much emphasis on evaluating the absolute values of mechanical properties. The absolute value of different properties might differ with the route of fabrication techniques.

A. *Thermal shock*

The response to thermal shock on Cu-5 vol% $Al_2O_3$ microcomposites and nanocomposites has been assessed from the flexural strength on the basis of the 3-point bend test. Fig 3(a) represents the ultimate flexural strength of the composite under various thermal shock exposures. The specimens were subjected to up-thermal shock (treated at 193 K first then followed by 353 K temperature) and down-thermal shock (treated initially at 353 K and then at 193 K temperature) for various exposure times, which have been elaborated in Table. 1.

Microcomposite: Fig. 3(a) shows the effect of thermal conditioning on ultimate flexural strength of thermally shocked (up-thermal and down-thermal shock) Cu-$Al_2O_3$ micro- and nano-composites. The trend of increase in flexural strength of microcomposites under subjection of thermal shock is time dependent which is quite clear from Fig. 3(a). There is an increment of 21.95% in flexural strength value for the up-thermal shock treatment in module-1, whereas down-thermal shock treatment increases the bending strength value by 29.69%, illustrated in Fig. 3(b). The fracture surface illustrated in Fig. 4(a) symbolizes the microcomposite tested at ambient conditions. The down-thermal shock (i.e. 353 K to 193 K) promoted the intimate physical bonding of reinforcement-matrix. The reason could be attributed to the following: at 353 K the matrix around the alumina particles expands and it imparts a compressive force on the alumina particle which could lead to particle fragmentation sometimes (Fig. 4(b)). [18] This also leads to better mechanical interlocking of the particle with the matrix. [19] Whereas on the contrary when cooling occurs from the processing temperature during fabrication of composite the matrix shrinks around the reinforcement particle rendering residual tensile stresses/strains in the matrix and compressive stresses/strains in the reinforcement. [20] The increase in flexural strength of down-thermally shocked microcomposite is slightly higher than in up-thermal shock because the prior conditioning effect predominates over the later shock (which means that the thermal shock from (353 K to 193 K) is manifested as prior conditioning at 353 K temperature followed by immediate exposure to 193 K temperature i.e. the specimen was conditioned at 353 K temperature for 1 hour and then immediately exposed to 193 K temperature. Thus the prior thermally conditioned specimen has experienced a thermal shock of 433 K temperature). The microcomposites when subjected to up-thermal shock, (i.e. 193 K to 353 K) the contraction of matrix (copper) takes place to a larger extent than the reinforcement particle (alumina). The reason for this could be the higher co-efficient of thermal expansion of copper (16.6 x$10^{-6}$ $k^{-1}$) than alumina (5.4 x $10^{-6}$ $k^{-1}$), which also aids in the presence of dislocation density at the interface. [21] This may lead to physical de-cohesion (Fig. 4(c)) at the particle/matrix conjunction which has been observed earlier by Ray et al. [10] in inorganic fiber/polymer composite. The dislocations present in the reinforcement proximity also get pinned down when the composite is subjected to 193 K. The pinning of dislocations strengthens the composite by resisting the plastic flow of the matrix. [22] The interfacial de-cohesion which had possibly occurred at ultra low temperature (Fig. 2(c)) could not be restored at 353 K on the same scale. The conditioning at 193 K for 60 minutes might have created a large interfacial mismatch at the interface which is manifested by the presence of differential co-efficient of thermal expansion between copper matrix and alumina reinforcement particle. [19]

Nanocomposite: The ultimate flexural strength value decreases for up-thermal shock whereas the strength values increases for down-thermal shock which is confirmed from Fig 5(a) & (b). In up-thermal shock the matrix in the vicinity of the particle contracts at 193 K temperature possibly leading to interfacial de-cohesion. The surface area of the nano particles being higher, the degree of de-union anticipated is also high rendering the interfacial bond weak. The above being a physical phenomenon cannot be reversed/restored at 353 K temperature, hence the ultimate flexural strength decreases. The surface deactivation of alumina nanoparticles which took place at 193 K was unlikely to be restored at 353 K. The adverse effect of prior thermal conditioning treatment on the composites is not being reversibly reinstated by the subsequent treatment; this could be a possible reason for a reduced value of flexural strength.

In down-thermal shock, the degree of physical contact of matrix and the reinforced nanoparticle increases at 353 K temperature. (The expansivity of copper matrix is much higher compared to that of alumina particle. So, the expansion of matrix onto the particle at +80˚C leads to enhanced gripping of alumina by the matrix. This enhanced proximity leads to mechanical strengthening of the interface which is reflected by the increased flexural strength values.) The improved integrity can also be attributed to the enhanced surface diffusivity of nanoparticles at high temperature. The physical integrity of copper and alumina has been shown by flexural strength. Nanoparticles have high surface energy leading to high surface diffusivity. At high temperature it is reasonably expected that the surface diffusivity of nanoparticles gets improved, and at sub-zero temperature surface diffusivity gets reduced, as diffusion is a temperature dependent phenomenon.

The ultimate flexural strength increases due to the high surface area of nanoparticles and hence lead to enhanced interaction of nanoparticles with the matrix at high temperature. Later the exposure at 193 K temperature could not induce damage on the same scale as the prior treatment at 353 K temperature. The fracture surface of the nanocomposite after down-thermal shock (Fig.

6(a)) show some tear ridges. Fig. 6(b) & (c) shows the load-displacement curve for micro- and nano-composites tested at ambient temperature.

The Al/AlN composites were prepared by squeeze casting, and were solution treated at 530˚C and quenched in water for 2 hours followed by aging for 10 hours at 160˚C. Thermal cycling led to increase in tensile strength, elastic limit and yield strength, and over all properties stability of the composites which is in accordance with our case, where flexural strength increases with thermal shock treatment. [23] Bhattacharya et al. have reported increase in microhardness and decrease in density (due to formation of voids) after thermal cycling in Al-SiC composites. [24] Cracks at the interface have been observed due to thermal strain in the composite in the clustered region of reinforcement. Our investigation also reports particle cracking after thermal shock treatment.

B. *Thermal conditioning*

Microcomposite: At 353 K temperature, the ultimate flexural strength of microcomposite decreases by 29.27% when compared to the untreated sample. The 3-point flexural strength values at various conditioning temperatures are illustrated in Fig. 7(a). As the expanding elastic matrix imparts a compressive force on the reinforcement particle, this results in particle fragmentation (abundantly visible in SEM micrograph Fig. 7(b)) which leads to composite softening. This differential expansion also leads to localized stresses and strain fields in the microcomposite. [25] The ultra low temperature conditioning at 193 K may render shrinkage of matrix which causes interfacial de-cohesion (Fig.7(c)) decreasing the flexural strength of microcomposite by 0.93%. At low temperature the degree of contraction of matrix is higher than the reinforcement particle shrinkage. As the probability of particle cracking decreases, this results in decrease of the detrimental softening effect. So, the decrease in ultimate flexural strength is less pronounced as compared to the 353 K conditioning.

Nanocomposite: At 353 K temperature, the ultimate flexural strength increases by 15.60 % in comparison to the untreated sample (Fig. 8). The nano alumina particles impede the dislocation motion leading to the dislocation pileups at the reinforcement particle-matrix boundary, which leads to strain hardening of the composite. The enhanced surface diffusivity of nanoparticles at high temperature could be another reason for proficient interfacial interaction and subsequent composite strengthening. On the contrary the ultra low temperature conditioning of the nanocomposite at 193 K decreases the composite strength by 26.31%. As the movement of dislocations is a temperature driven phenomena, at low temperature the movement of dislocations get arrested which restricts the dislocation pile up consequently decreasing the density of dislocation forest. [26] At low temperature the surface diffusivity of alumina nanoparticles is quite likely to get lowered in comparison to enhanced diffusivity at high temperature, hence the thermal conditioning effect in nanocomposites is pronounced.

C. *High temperature 3-point bend test*

Microcomposite: At 373 K temperature the copper matrix expands resulting in tensile and compressive stress in the matrix and reinforcement respectively enhancing the mechanical interlocking between matrix and reinforcement. [27] This leads to the direct strengthening of the composite and as a result the ultimate flexural strength of the composite increases by 18.12% from the ambient test value (illustrated in Fig. 9(a)).

At 523 K temperature the ultimate flexural strength values decreases by 20.56% from the room temperature test value. At higher operating temperatures plasticity of the composite increases due to (1) dislocation annihilation (2) activation of dislocation motion by different mechanism other than glide (3) relaxation of internal stress at the matrix-particle front, (4) enhancement of dislocation recovery at the interface. [28] All the above stated phenomena decrease the strain hardening exponent of the composite leading to decrease in ultimate flexural strength of the composite. The fracture surface (revealed from Fig. 9(b) & (c)) illustrates ample number of particle pull-out of alumina particles possibly due to matrix softening. [12]

Nanocomposite: With increase in operating temperature the strength of the Cu-$Al_2O_3$ nanocomposite decreases by 1.08 % at 373 K and 20.05% at 523 K temperature (Fig. 10(a)). With increase in operating temperature the strengthening mechanisms that operate at low temperature get relaxed which decrease in the strength of the composite. [27] The fractography studies reveal that the fracture mode of the nanocomposite at 373 K is of ductile type (Fig. 10(b)) whereas the failure characteristic features of the composite at 523 K temperature can be characterized by dimple markings (matrix softening at elevated temperature). The nanocomposites do not show pronounced dimples in ambient conditions of testing, whereas at high temperature of testing presence of dimples suggest matrix softening to a certain extent.

Uematsu et al. subjected Al-SiC composites to high temperature testing, showing decrease in tensile strength with increase in testing temperature such as 150˚C and 250˚C. [29] The fracture surface indicated particle fracture and particle/matrix crack intiation due to softening of matrix at high temperature which is in agreement with our results.

D. *Comparison of behavior of differently treated composites*

The present investigation has largely emphasized the study of damage and/or development of flexural properties under the condition of up-thermal and down-thermal shocks. The experiment has further focused on the variation of flexural behavior by the imposition of thermal conditioning and high temperature exposure on micro- and nano-particle embedded Cu based composites.

*Thermal shock*

The composites are exposed to temperature gradient of 433 K (from 353 K to 193 K i.e. down-thermal shock, and 193 K to 353 K temperature i.e. up-thermal shock). This thermal shock experiment revealed that the flexural strength of microcomposites increases after down- as well up-thermal shock treatments. Whereas, for nanocomposites there is an incremental improvement in flexural strength after down-thermal shock conditioning and the decrease of its value has been observed after up-thermal shock treatment.

The larger surface/volume ratio of alumina nanoparticles in comparison to microparticles implies more surface area contact with the copper matrix for the former. The implications of thermal treatments for both the up-and down-thermal shocks are likely to affect more particle/matrix interfaces in nanoparticles embedded copper matrix system. Thus it may reasonably be proposed that the differential co-efficient thermal expansion of copper and alumina may manifest larger amount of interfacial damage in nanosystem, because of an exposure to a temperature gradient in a short span of time. That is why more degradation has been observed in nanocomposites.

The enhancement of property has also been observed in few cases for both the systems. This may be manifested by an improvement in mechanical interlocking factor. This improvement in interlocking factor may sometimes get nullified and/or diminished during exposure in an opposite direction of thermal cycling. These contradictory and inconclusive natures of behaviour might be attributed to the generation of opposite and complex residual stresses during exposure to high to low and low to high temperature cycles.

*Thermal conditioning*

The specimens had experienced a temperature of 353 K and 193 K separately which were conditioned at that temperature for an hour. The microcomposites reflect a decrease in flexural strength after 353 K temperature conditioning. Nanocomposites show an increase in flexural value at 353 K temperature, and decrease at 193 K temperature conditioning respectively.

The reduction in mechanical property during cryogenic conditioning may be attributed by the de-cohesion between alumina particle and copper matrix in nanocomposites. This might have been manifested because of large difference in co-efficient of thermal contraction between particle and matrix.

The high temperature exposure of nanocomposites has demonstrated an increase in mechanical property. It may be reasonably assumed that differential expansion may be leading to better particle matrix registry and thereby closer proximity is ensured.

The lower surface area/volume ratio of microparticle in comparison to the nanoparticle embedded copper matrix has shown no significant changes for cryogenic temperature conditioning and a decline in flexural strength at high temperature conditioning. The forced close intimacy of the alumina microparticle and matrix at high temperature leads to development of residual stresses at the particle/matrix interfacial region. This non-uniform distribution of residual stresses may decrease the threshold value of crack nucleation and propagation along the particle/matrix interface region. The differential co-efficient of thermal expansion between alumina particle and copper matrix has lesser impact and implications either in advancement or declination of mechanical property in micro-alumina reinforced copper composites.

*High temperature flexural test*

The composites were tested at 373K and 523 K temperatures, and these temperatures were maintained constant throughout the experiment. The increase in flexural strength at 373 K and decrease in flexural strength at 523 K temperatures reflect the poor high temperature sustainability of microcomposites. The nancomposites also reveal low flexural strength at 523 K.

The increase in flexural strength at 373K is due to the effective gripping of alumina particle and copper matrix due to relatively higher expansion of copper than alumina. The residual stress development triggers the forced close registry between particle and matrix imposed by high temperature (523K) conditioning. This may not be conducive for the generated residual stresses to be distributed properly and uniformly. These accumulated non-uniform residual stresses may decrease the threshold value of crack nucleation and propagation along the particle/matrix interface region.

## IV. CONCLUSIONS

Copper with 5 vol. % alumina micro- and nano-composites were fabricated by conventional powder metallurgy route. Both up- and down-thermal shock treatments enhance the ultimate flexural strength of Cu-$Al_2O_3$ microcomposites. The variation of flexural strength is contradictory and far from comprehensive conclusion in Cu-$Al_2O_3$ nanocomposites. Thermal conditioning at 353 K temperature improves the ultimate flexural strength of Cu-$Al_2O_3$ nanocomposites. At high operating temperatures (i.e. 523 K) the ultimate flexural strength of both Cu-$Al_2O_3$ micro- and nano-composites decreases. SEM micrographs reveal ductile mode of fracture for both micro- and nano-composites. Ductile fracture characteristics have been observed predominantly for Cu-$Al_2O_3$ microcomposites. Fracture characters visible in Cu-$Al_2O_3$ nanocomposites indicate ductile mode of failure. The nanoparticles in nanocomposites have higher surface area as compared to microparticles in microcomposites. So, the thermal shock induced stress in nanocomposites is more visible in terms of degradation and enhancement of flexural strength. The development and detoriation of physical integrity of composite is predominant in nanocomposites than microcomposites**.** Noticeable differences in the flexural strength and response to thermal exposures of the micro- and nano-composites have been observed, and have been explained in terms of difference in their fracture surface microstructures.

## ACKNOWLEDGMENT

The authors would like to thank the National Institute of Technology (NIT), Rourkela for providing the necessary financial and infrastructural supports.

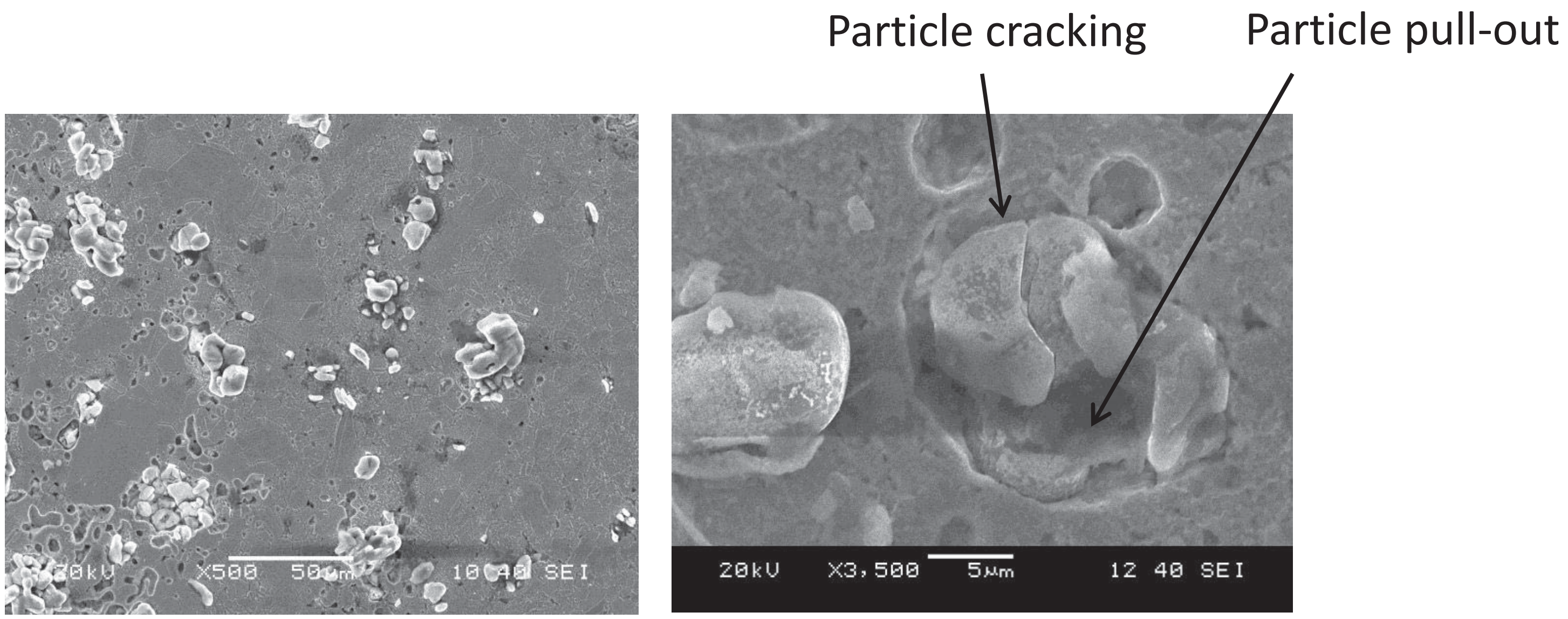

Fig. 1— SEM micrographs of Cu-5% $Al_2O_3$ microcomposite (left) before any thermal treatment (right) after down-thermal shock (right figure is currently under consideration for the proceedings of International Conference on Recent Advances in Composite Materials 2013)

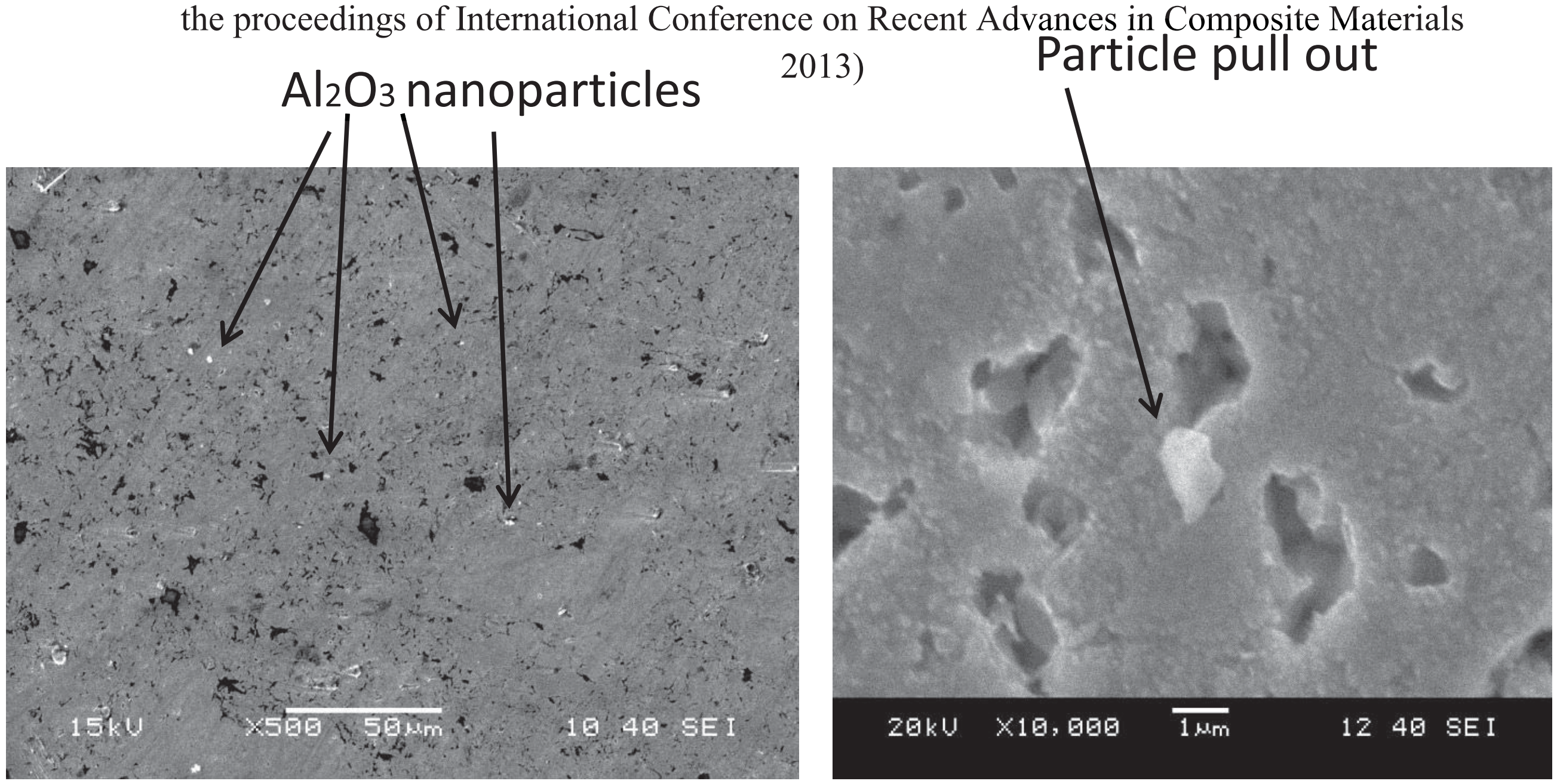

Fig. 2— SEM micrographs of Cu-5% $Al_2O_3$ nanocomposite (left) before any thermal treatment (right) after down-thermal shock (right figure currently under consideration for the proceedings of International Conference on Recent Advances in Composite Materials 2013)

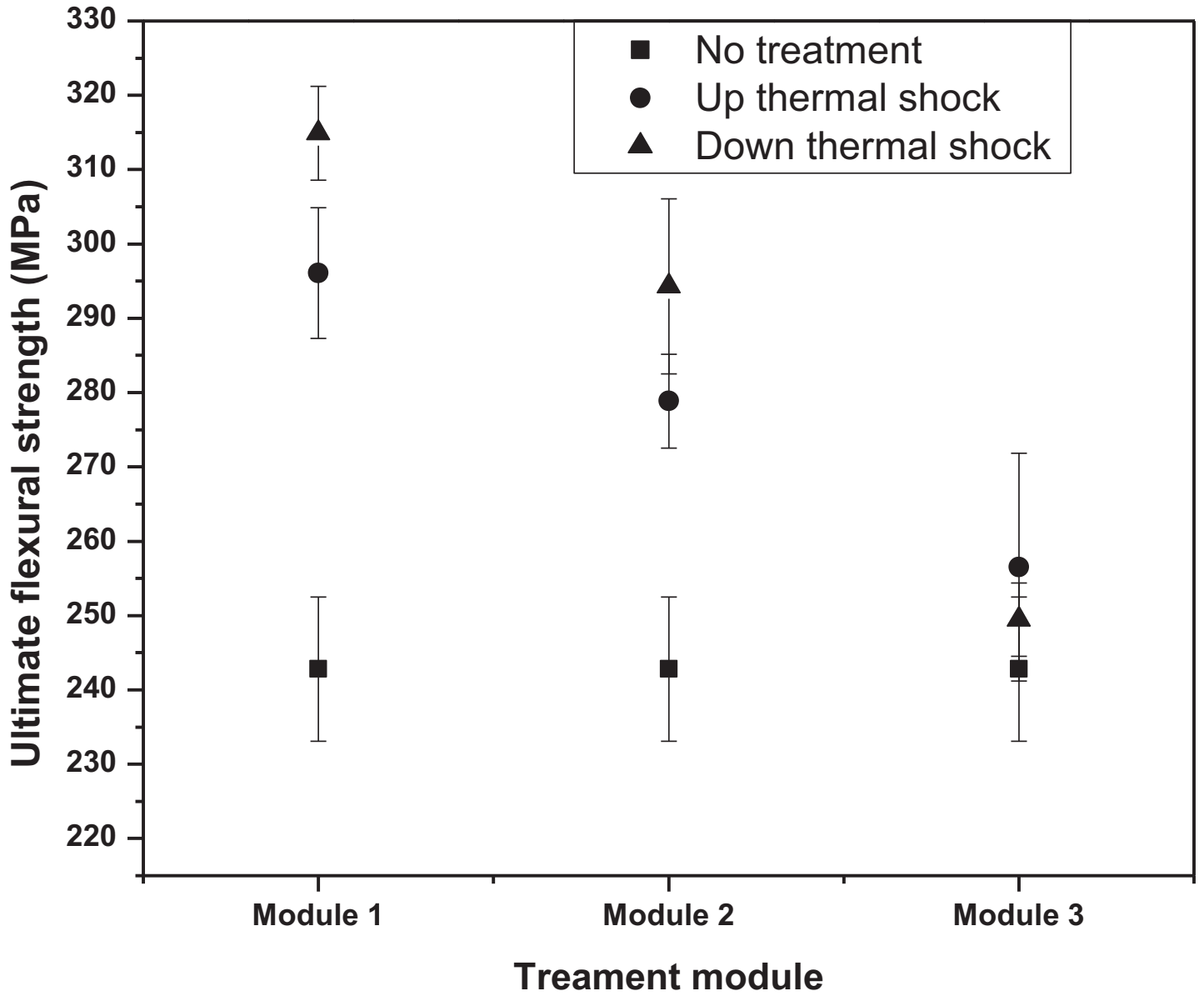

Fig. 3(a)—Comparison of ultimate flexural stress (MPa) value for up-thermal shock and down-thermal shock of microcomposites treated in different treatment modules.

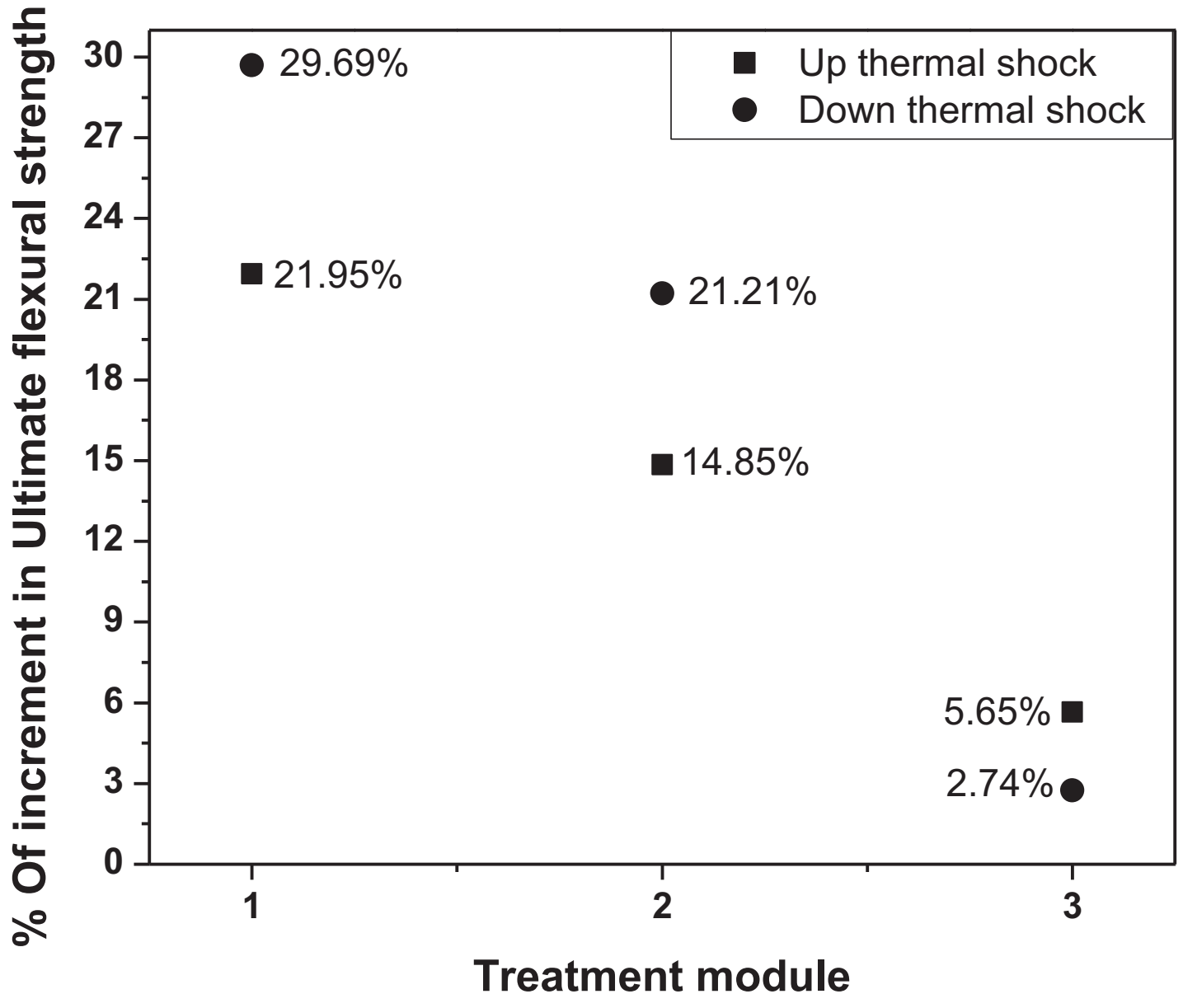

Fig. 3(b)—Percentage of increment in ultimate flexural stress (MPa) value for up-thermal shock and down-thermal shock with respect to ambient values of microcomposites treated in different treatment modules.

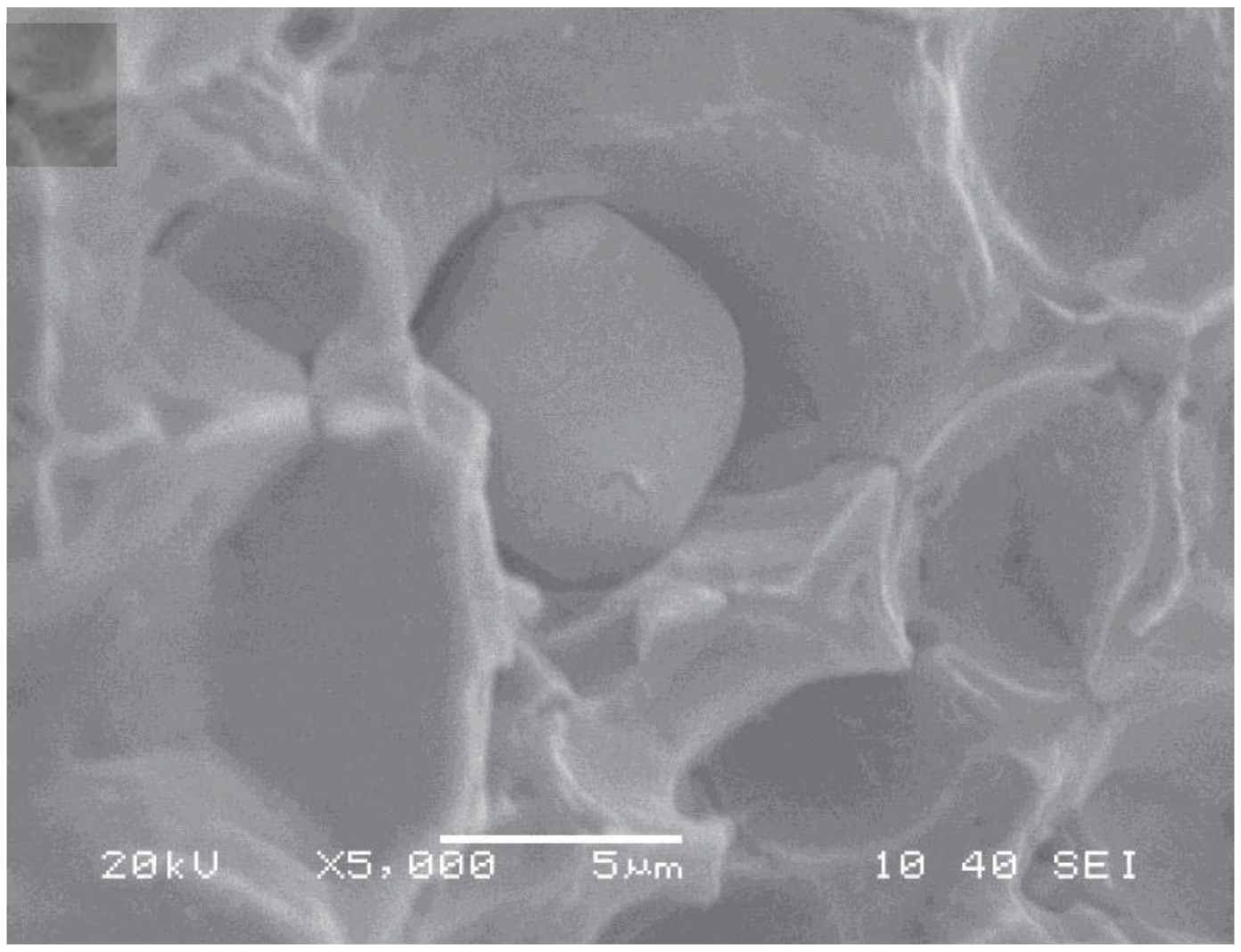

Fig. 4(a)

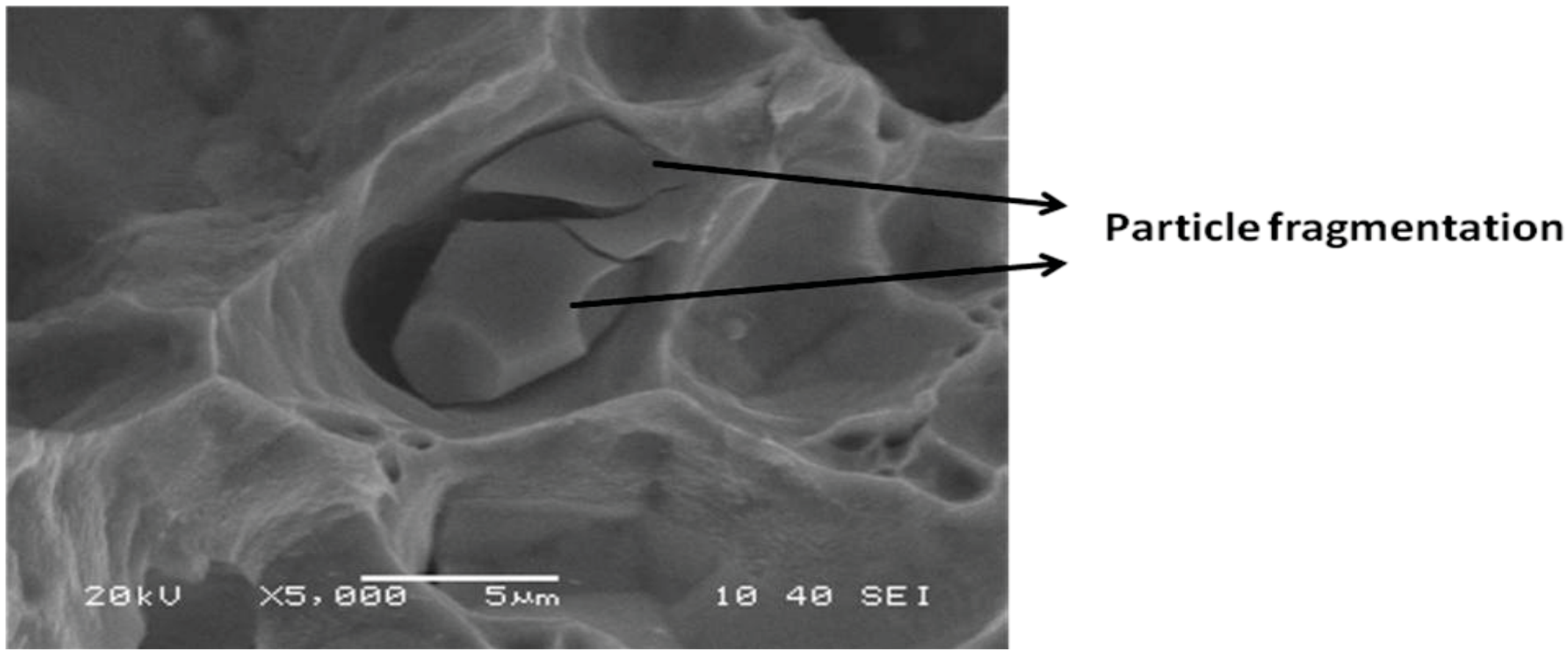

Fig. 4(b)

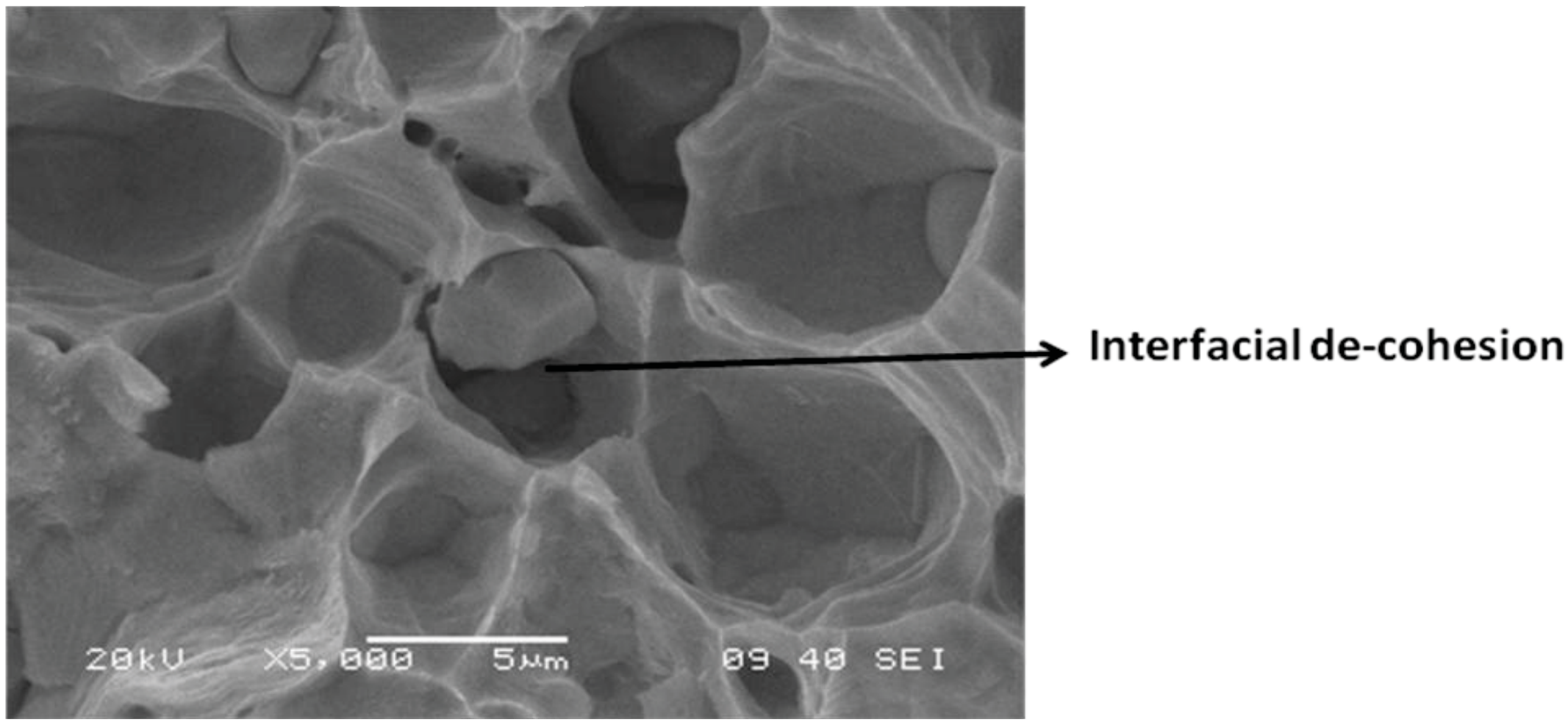

Fig. 4(c)

Fig. 4—SEM micrograph illustrating the thermally shocked 3-point bend test specimen fracture surface of microcomposites (a) no treatment (b) down-thermal shock (c) up-thermal shock.

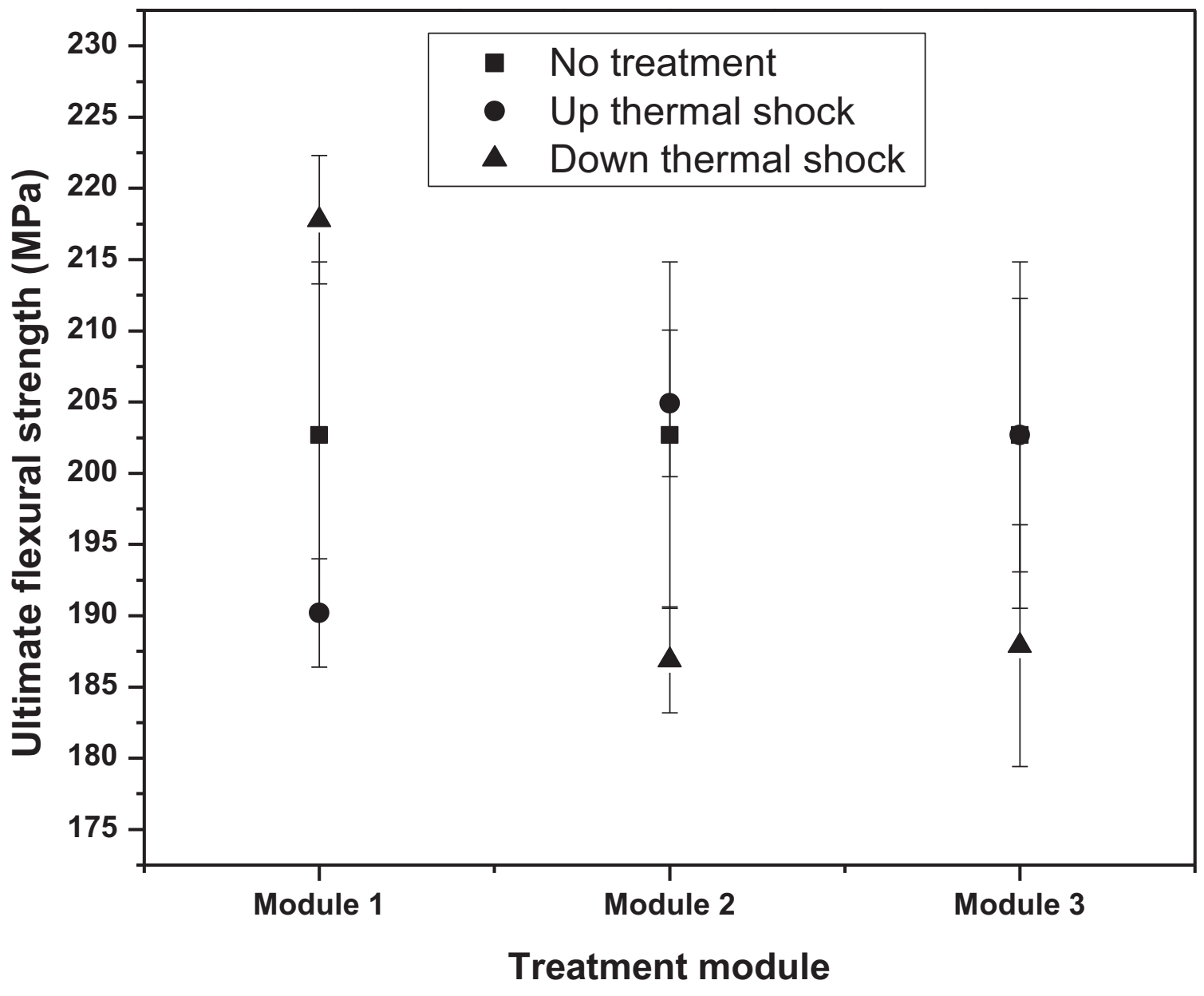

Fig. 5(a)— Comparison of ultimate flexural stress (MPa) value for up-thermal shock and down-thermal shock of nanocomposites treated in different treatment modules.

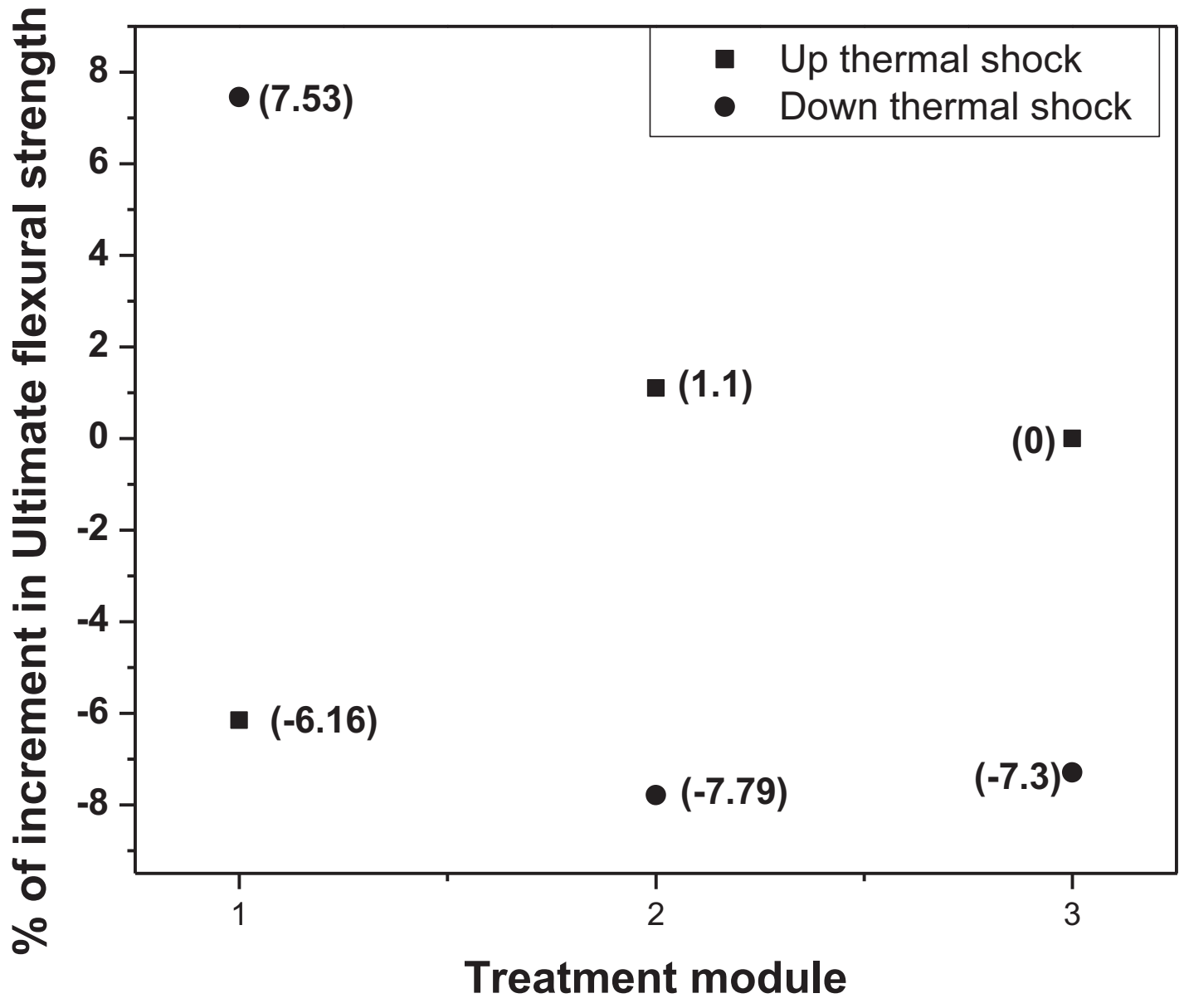

Fig. 5(b)—Percentage of increment in ultimate flexural stress (MPa) value for up-thermal shock and down-thermal shock with respect to ambient values of nanocomposites treated in different treatment module.

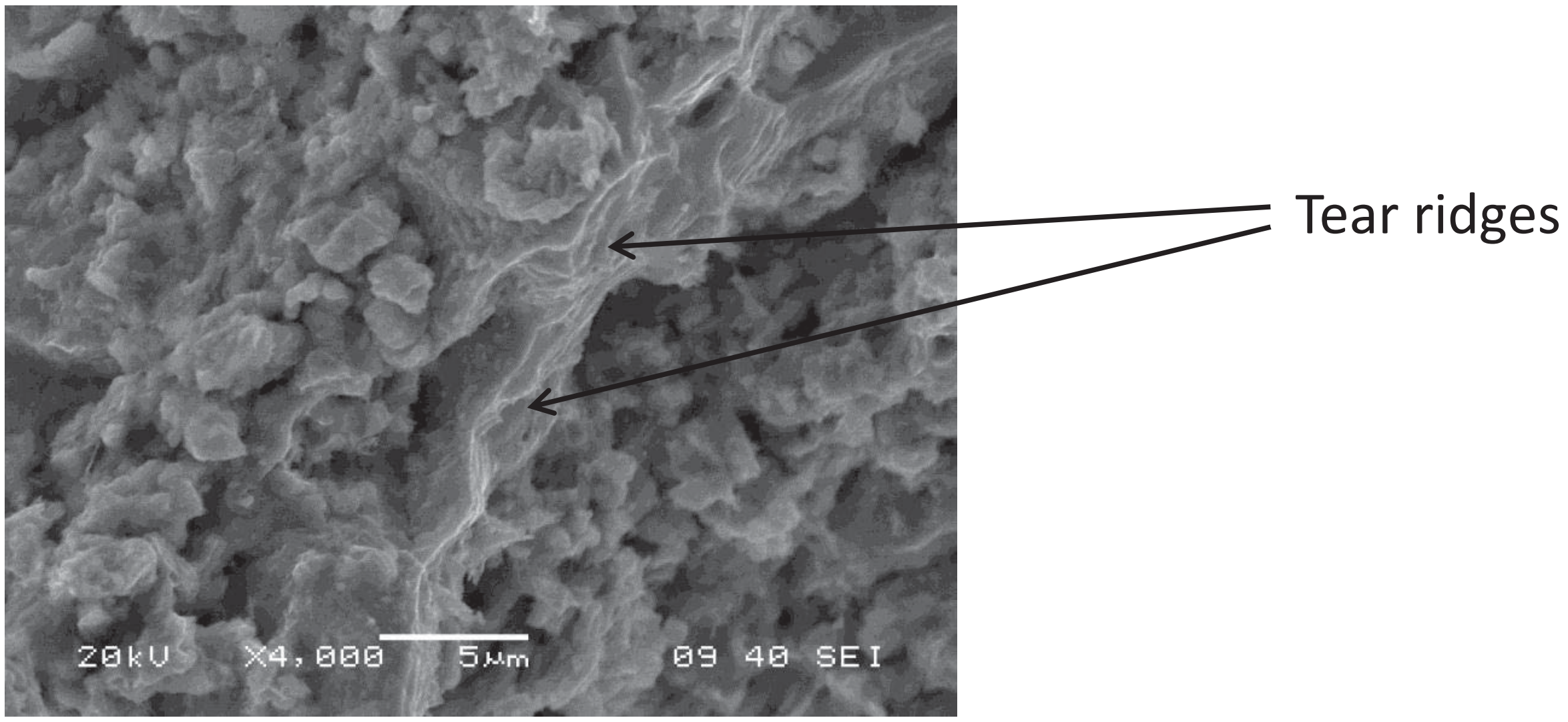

Fig: 6(a)—SEM micrographs illustrating the thermally shocked 3-point bend test specimen fracture surface of nanocomposites down-thermal shock.

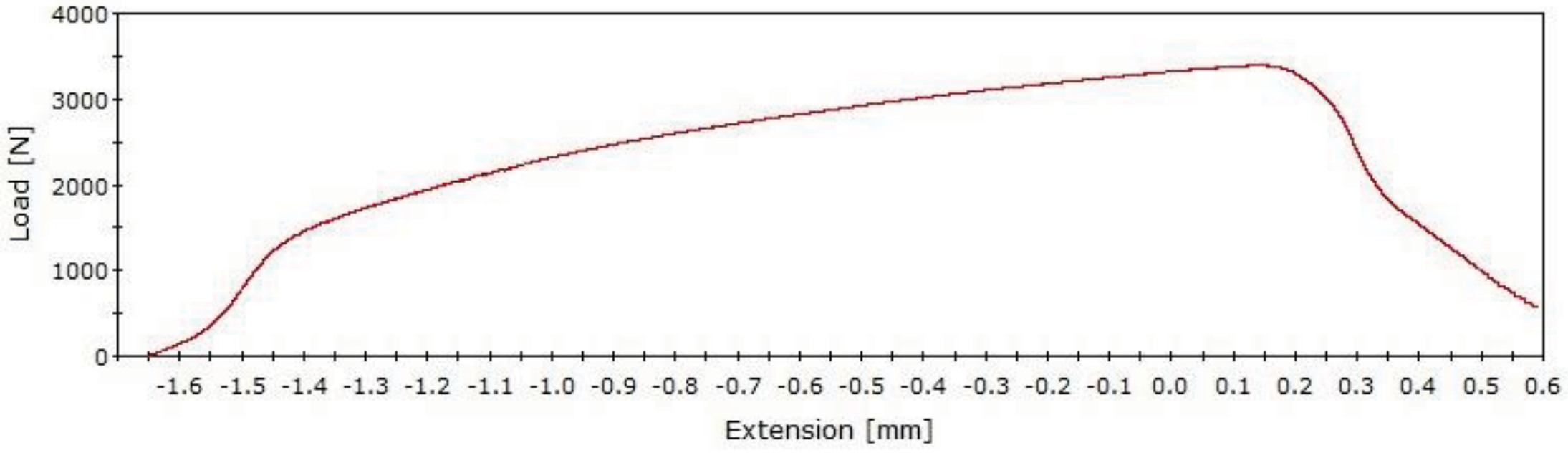

Fig. 6(b)—Load-displacement curve for Cu-5% $Al_2O_3$ microcomposite tested at ambient conditions.

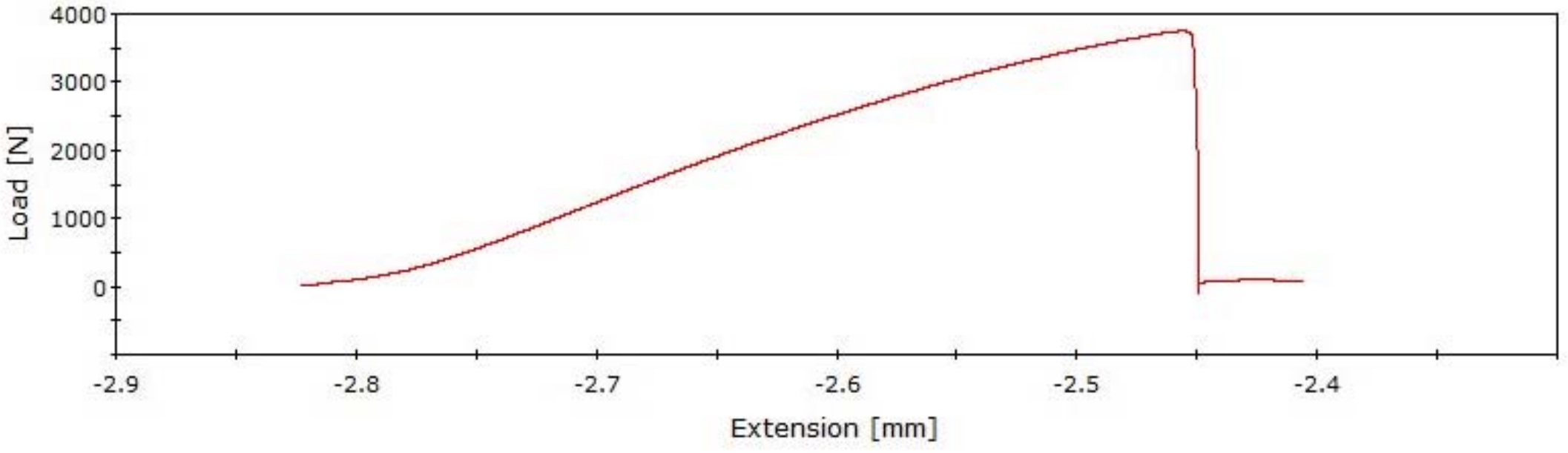

Fig. 6(c)—Load-displacement curve for Cu-5% $Al_2O_3$ nanocomposite tested at ambient conditions.

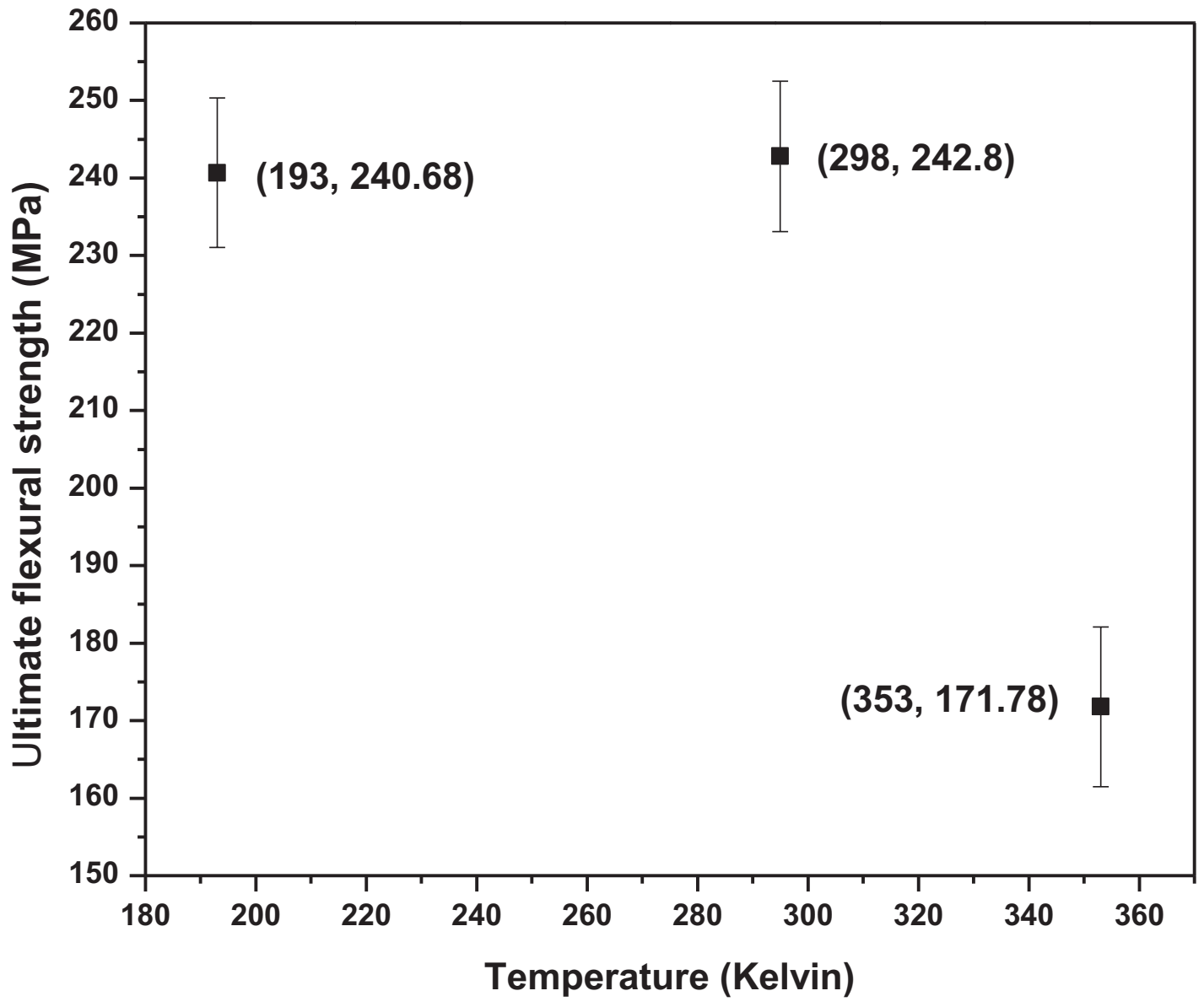

Fig. 7(a)— Plot for ultimate flexural stress (MPa) vs. thermal conditioning temperature for microcomposites.

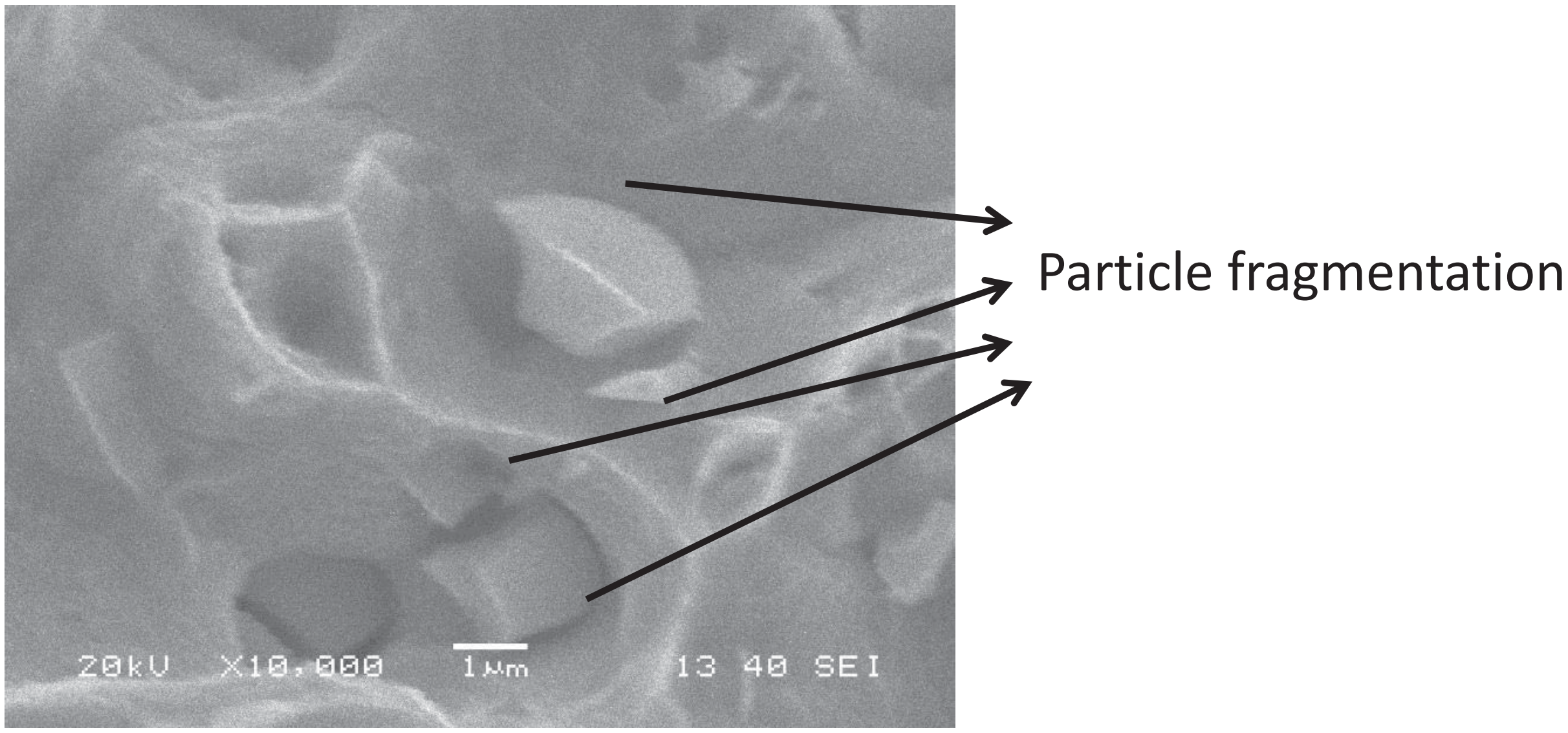

Fig. 7(b)

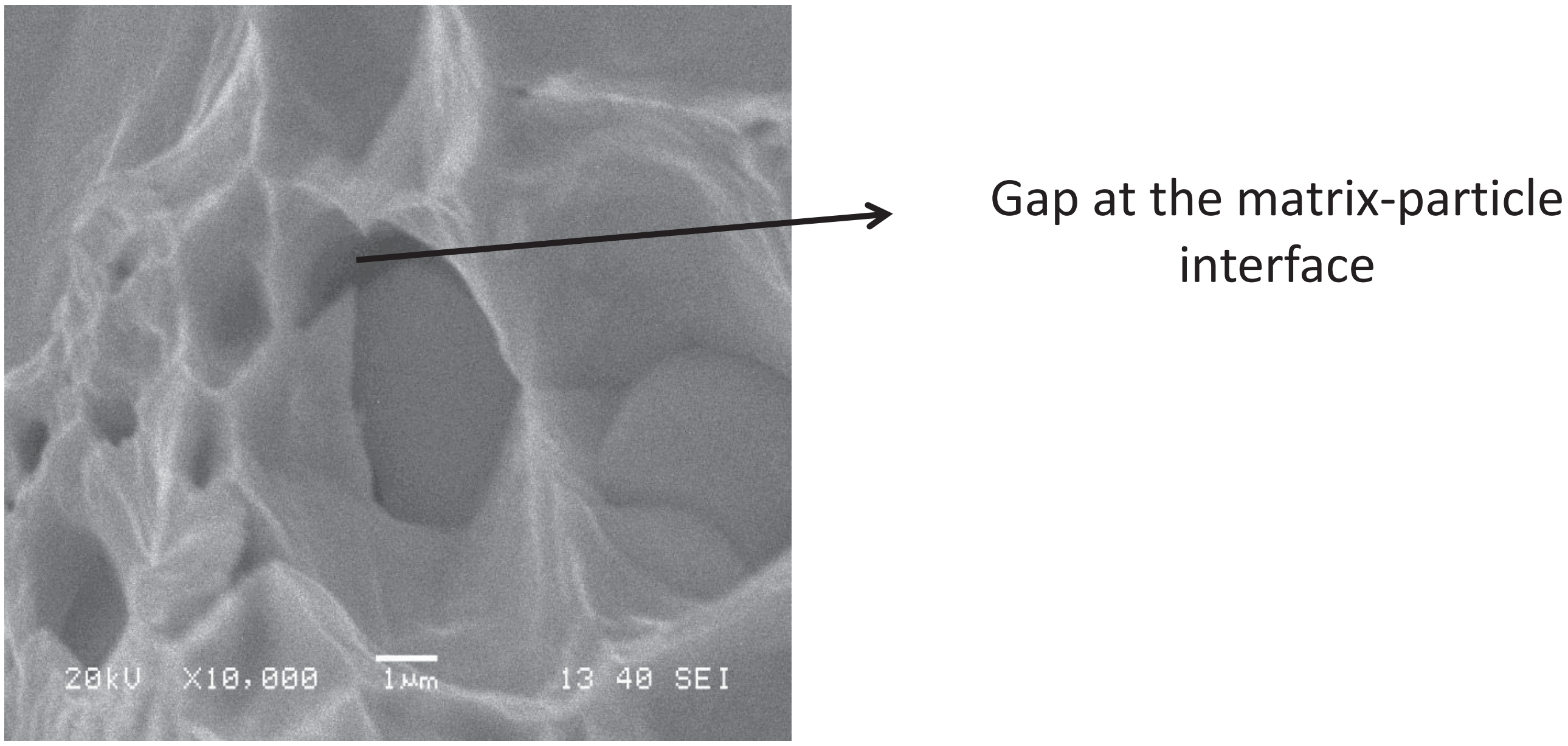

Fig. 7(c)

Fig. 7—SEM micrographs illustrating the thermal conditioned 3-point bend test specimen fracture surface of microcomposites (b) conditioning at 353 K (80˚C), (c) conditioning at 193 K (-80˚C).

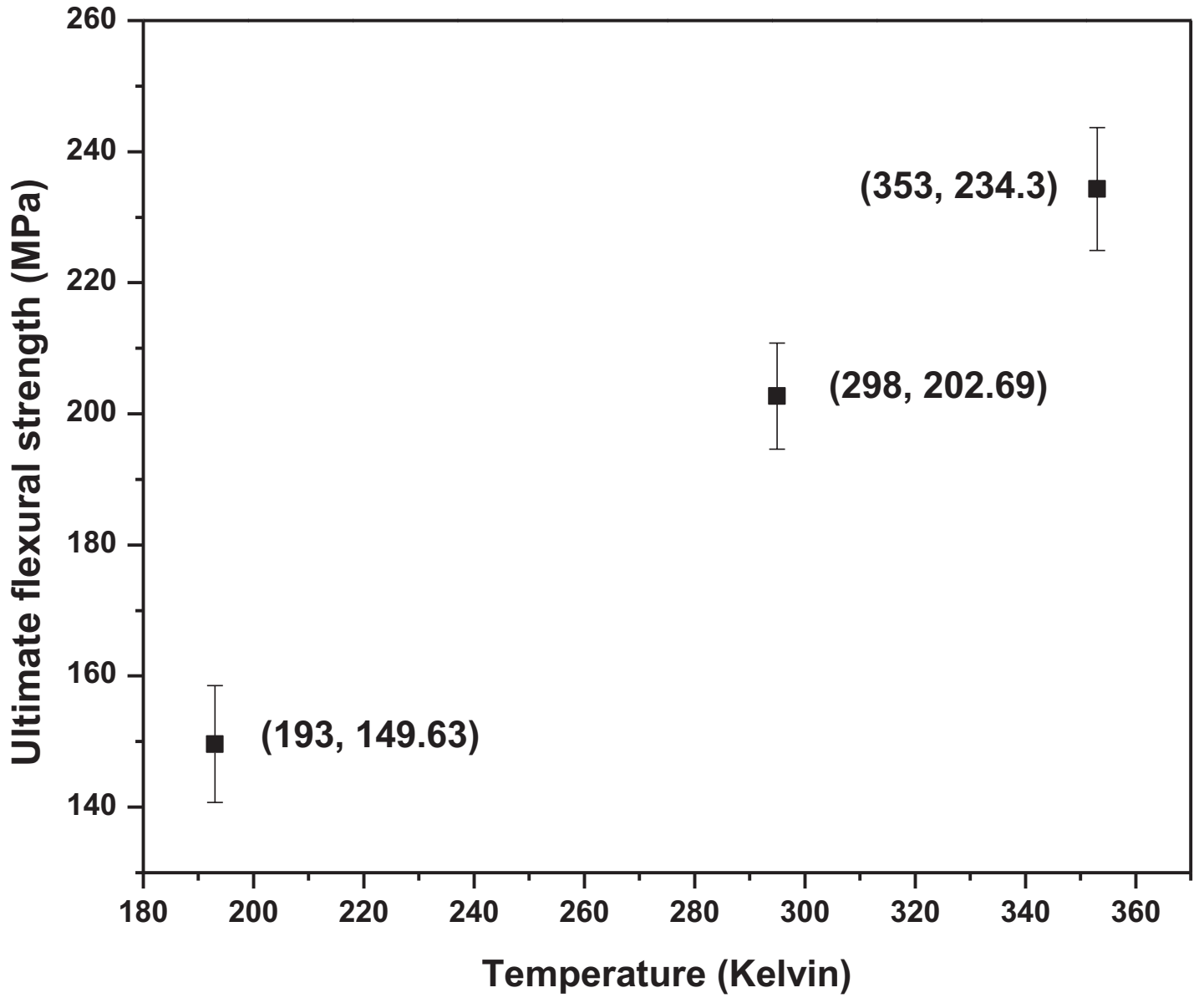

Fig. 8—Plot for ultimate flexural stress (MPa) vs. thermal conditioning temperature for nanocomposites

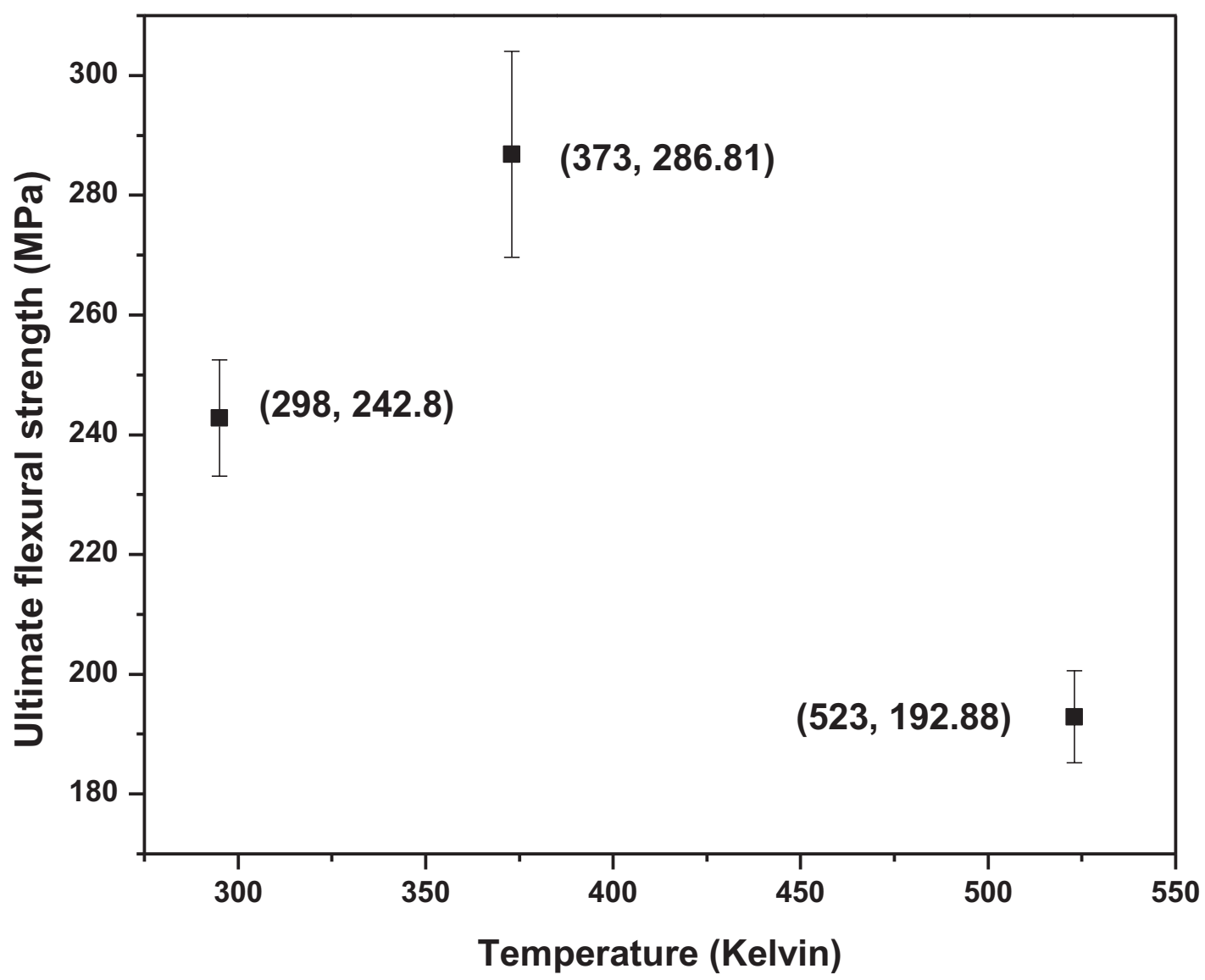

Fig. 9(a)— Plot for ultimate flexural stress (MPa) vs. operating temperature for microcomposites in high temperature 3-point bend test.

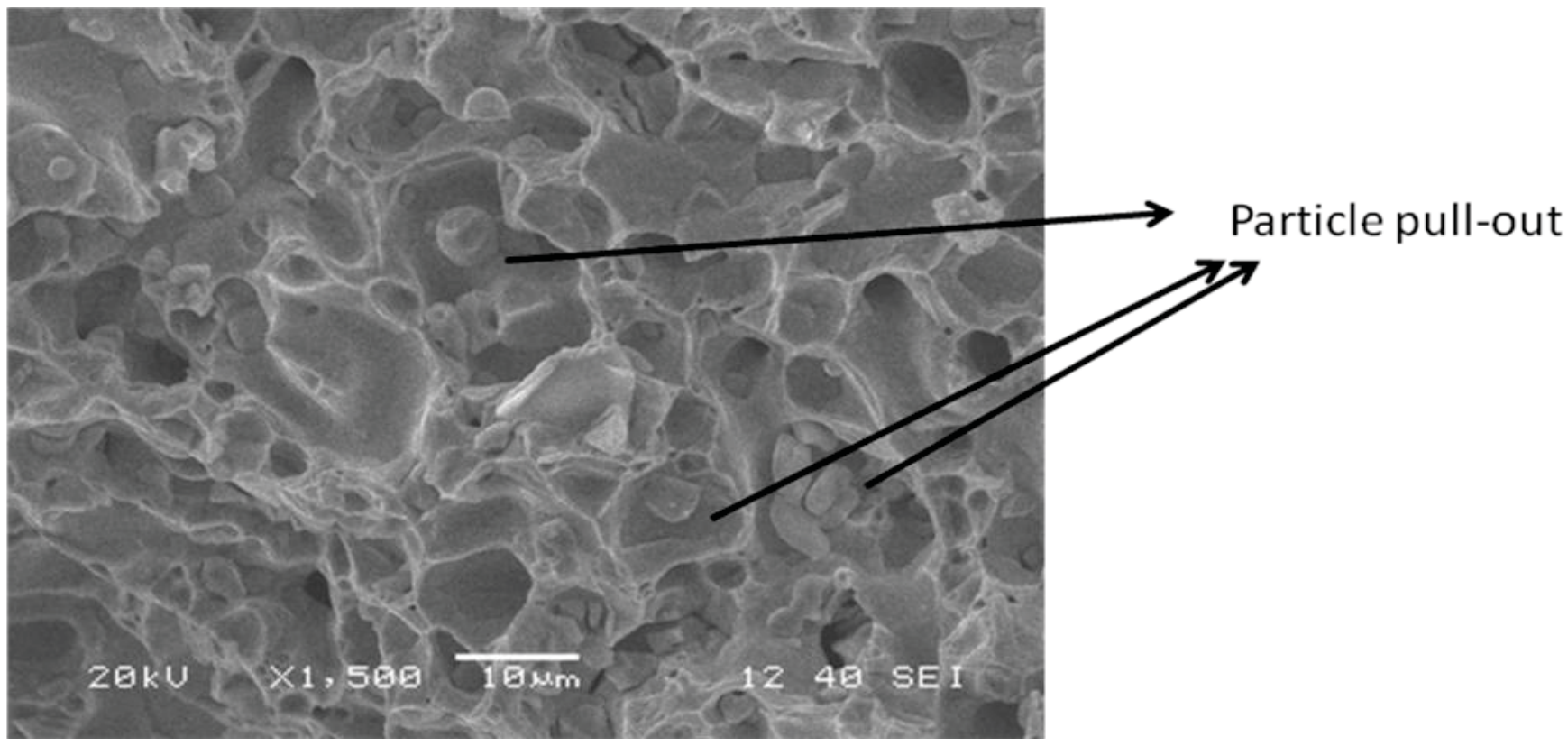

Fig. 9(b)

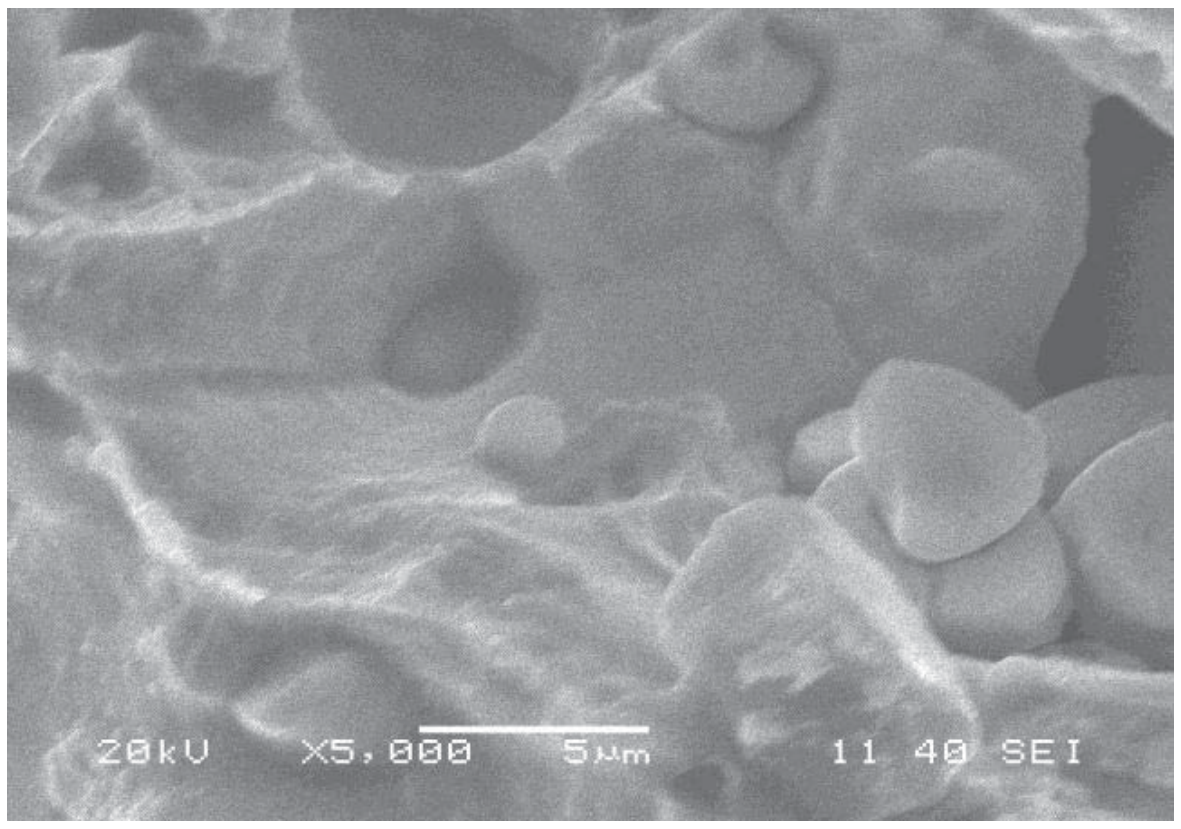

Fig. 9(c)

Fig. 9—Fractography of the high temperature test (b) at 373 K (100˚C) and (c) at 523 K (250˚C) of microcomposites.

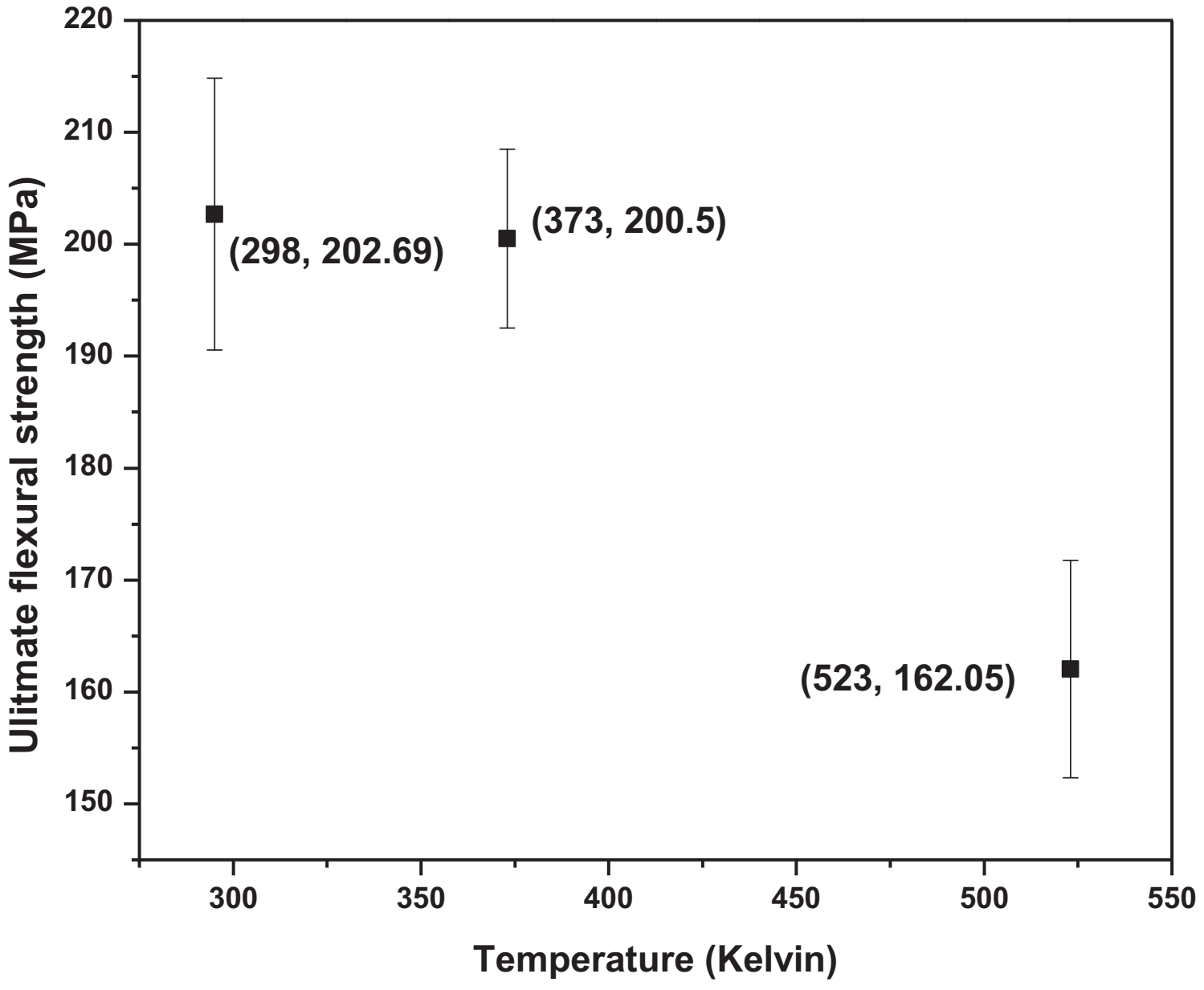

Fig. 10(a)— Plot for ultimate flexural stress (MPa) vs. operating temperature for nanocomposites in in-situ high temperature 3-point bend test.

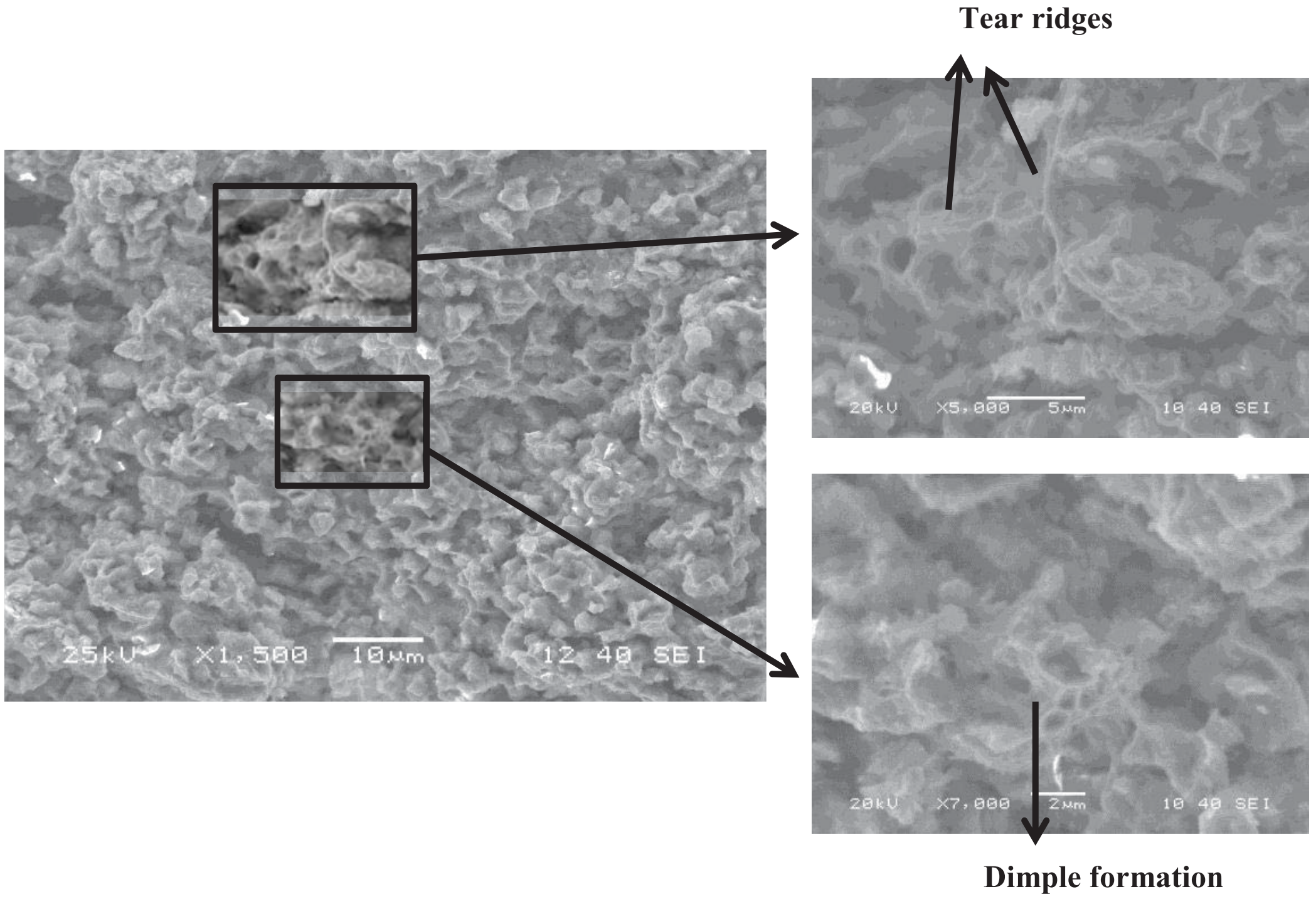

Fig. 10(b)— Fractography of the high temperature 3-point bend test at 523 K (250˚C) of nanocomposites.

Table 1: Detailed description of types of thermal shock treatment in terms of conditions followed and time period of exposure

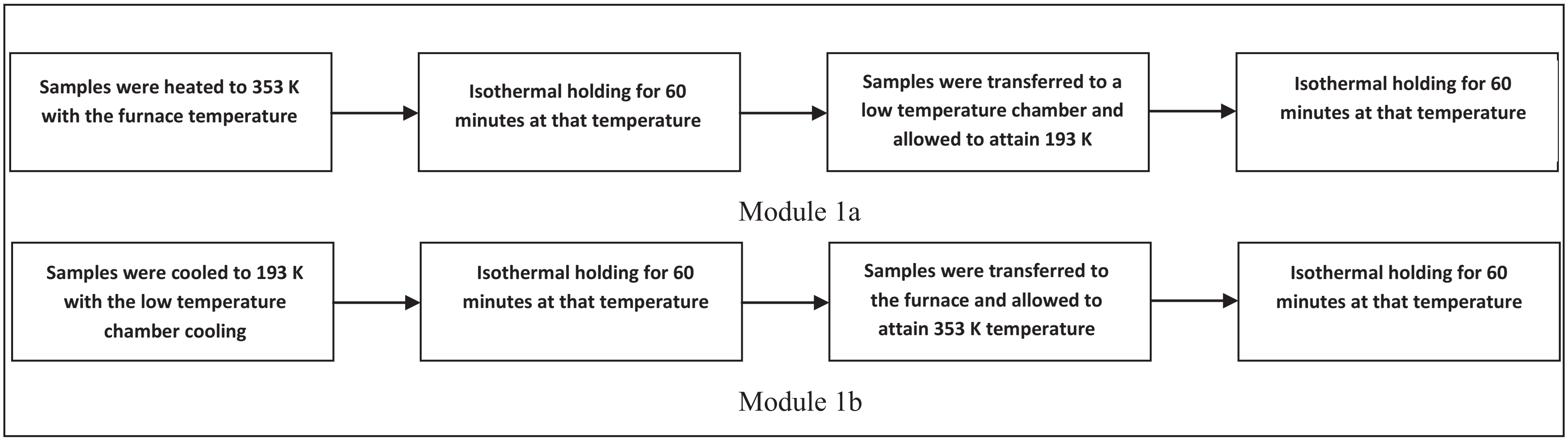

Module 1— (a): Down-thermal shock (b): Up-thermal shock

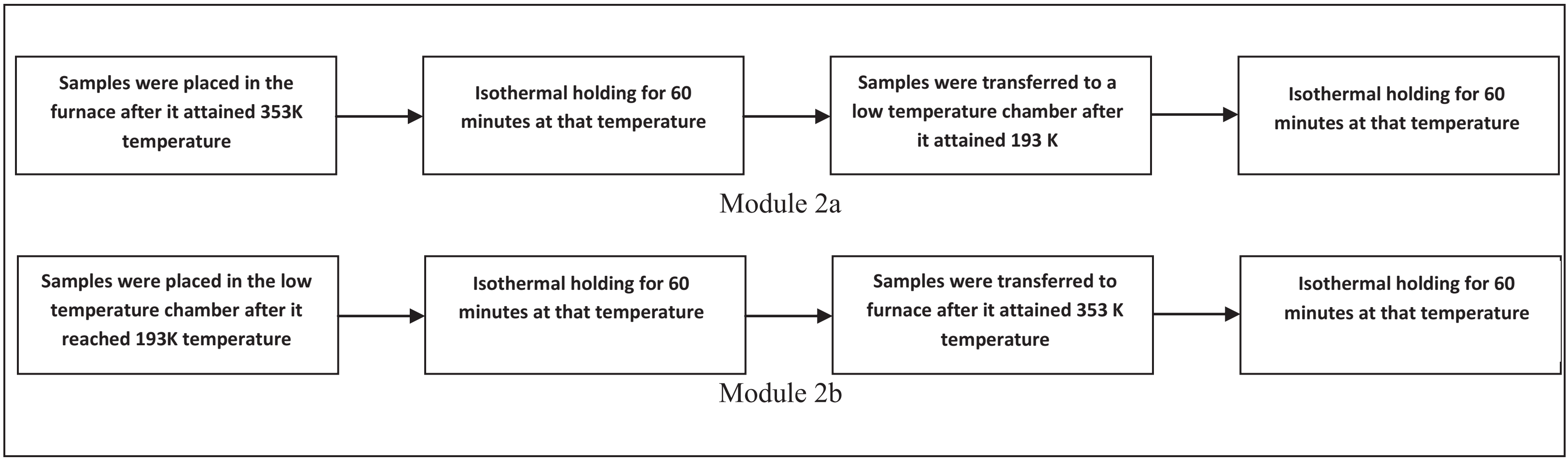

Module 2— (a): Down-thermal shock (b): Up-thermal shock

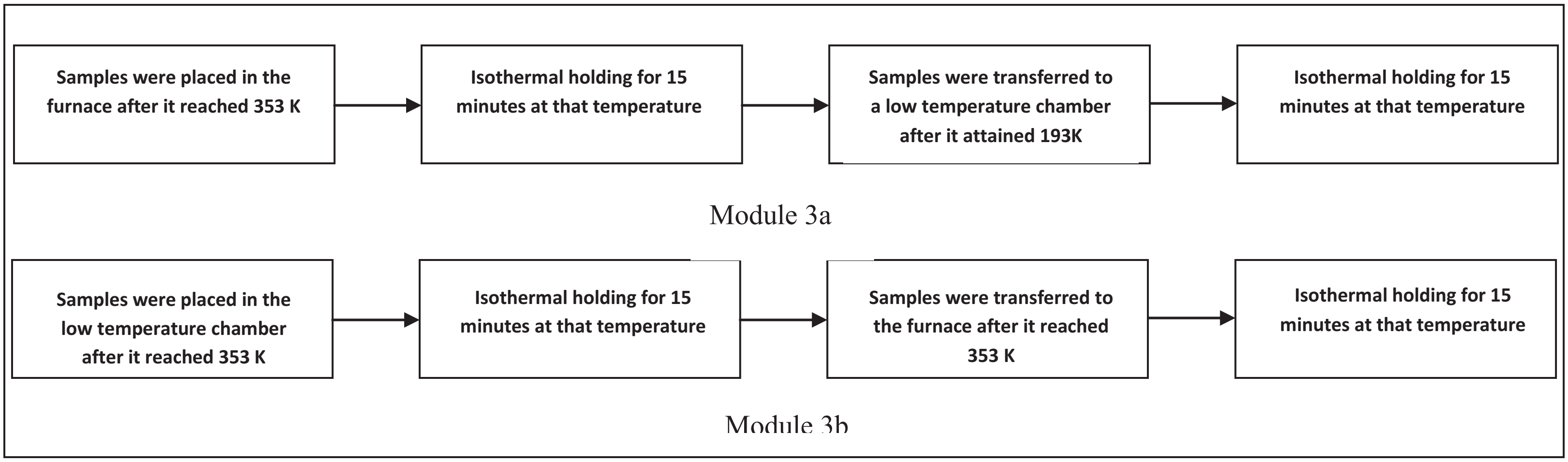

Module 3— (a): Down-thermal shock (b): Up-thermal shock